\documentstyle[preprint,revtex]{aps}
\begin{document}
\draft
\preprint{ to be published in Phys. Rev. C}
\begin{title}
\begin{center}
{General Behavior of Double Beta Decay Amplitudes in the Quasiparticle
Random Phase Approximation}
\end{center}
\end{title}
\author{F. Krmpoti\'{c}}
\begin{instit}
Departamento de F\'\i sica, Facultad de Ciencias Exactas,\\Universidad Nacional
 de La Plata, C. C. 67, 1900 La Plata, Argentina
\end{instit}
\begin{abstract}
Simple formulae for the $0^+\rightarrow 0^+$ double beta decay matrix
elements, as a function of the particle-particle strength $g^{pp}$, have been
designed within the quasiparticle random phase approximation. The $2\nu$
amplitude is a bilinear function of $g^{pp}$, and all $0\nu$ moments behave as
ratios of a linear function and the square root of another linear function of
$g^{pp}$. It is suggested that these results are of general validity and that
any modifications of the nuclear hamiltonian or the configuration space
cannot lead to a different functional dependence.
\end{abstract}
\narrowtext
The neutrinoless double beta decay ($0\nu \beta \beta$) is very interesting
for several reasons.
In the first place, this  decay mode is viable only when the neutrino is a
massive Majorana particle. As such, it constitutes a critical touchstone
for various gauge models that go beyond the standard $SU(2)_L\times U(1)$
gauge model of electoweak interactions.
Secondly, the neutrinos with nonzero masses have many interesting consequences
for the history of the early universe, in the evolution of stellar objects,
and the supernovae astrophysics.
Thirdly, besides the issue of $m_{\nu}\neq 0$, there are other open questions
in neutrino physics the answers to which depend on  $0\nu\beta \beta$ decay,
such as: Why does nature favor only left-handed currents? Does the majoron
exist?  Yet, we shall not understand the $0\nu\beta \beta$  decay unless we
understand the two neutrino double  beta decay ($2\nu\beta \beta$). The last
one is the rarest process observed so far in nature and offers a unique
opportunity for testing the nuclear physics techniques for half-lives $\agt
10^{20}$ years.  Thus, the comprehension of the $\beta \beta$ transition
mechanism cannot but help advance knowledge of physics in general.

In recent years the quasiparticle random phase approximation (QRPA) has been
the most popular method to deal with the problem of $0^+\rightarrow 0^+$ double
beta decay \cite{vog,eng,fes,kla,mut,tom,krm1,krm2,krm3}.
Within this model the $\beta \beta$-decay amplitudes are extremely sensitive
to the interaction parameter in the particle-particle (PP) channel,
usually denoted by $g^{pp}$.
Independently of the nucleus that decays, of the residual interaction
that is used, and of the configuration space that is employed, all the
 QRPA calculations done so far exhibit the following general features.

(i) Close to the "natural" value for $g^{pp}$ ($g^{pp}\cong1$)
the $2\nu \beta \beta$ moments have first a zero and latter on a pole at which
the QRPA collapses.

(ii) The  zeros and  poles of the $0\nu \beta \beta$ moments for the virtual
states with spin and parity $J^{\pi}=1^{+}$ are strongly correlated with the
zeros and poles of the $2\nu \beta \beta$ moments.

(iii) The  $0\nu \beta \beta$ moments of multipolarity $J^{\pi}\neq 0^+,1^+$
also possess zeros and poles but at significantly larger values of $g^{pp}$.

(iv) As a function of $g^{pp}$, both the $2\nu \beta \beta$ and
$0\nu \beta \beta$ moments always present similar shapes.

Fig.\ \ref{fig1} illustrates the behaviour of the $0^+\rightarrow 0^+$
$\beta \beta$ matrix elements for several nuclei. In the upper panel
the $2\nu \beta \beta$ moments (${\cal M}_{2\nu}$) are shown.
The other two panels contain the $0\nu \beta \beta$ moments of multipolarity
$J^{\pi}=1^{+}$ (${\cal M}_{0\nu}(J^{\pi}=1^{+})$) and total
$0\nu \beta \beta$ moments (${\cal M}_{0\nu}$), induced by the neutrino
mass mechanism. These results have been obtained with a $\delta $ force, using
standard parametrization presented elsewhere \cite{krm4}.
Instead of the parameter $g^{pp}$, I use here the ratio  between the triplet
and singlet coupling strengths  in the PP channel, i.e.,
$t={\it v}_t/{\it v}_s$. Calculations with finite range interactions yield
similar results \cite{fes,kla,mut,tom}.

More that once \cite{krm1,krm2,krm3} we have pointed out
that the $\beta \beta$ amplitudes go to zero within the QRPA because of
the restoration of both the isospin and SU(4) symmetries.
We have also suggested a physical criterion for fixing the PP coupling
strength based on the maximal restoration of the SU(4) symmetry ($t=t_{sym}$).
Yet, the general characteristics mentioned above suggest the existence of some
additional regularities, and the present concern reflects upon a global
understanding of the $\beta \beta$ transition mechanism within the QRPA.
Only in this way one can  get a full control of the calculations,
which is one of the prerequisites for a  reliable estimate of the nuclear
matrix elements.

To begin with, I resort to the single mode model  (SMM) description
\cite{krm3} of the $\beta \beta$-decays in the
$^{48}Ca \rightarrow\, ^{48}Ti$ and $^{100}Mo \rightarrow\, ^{100}Ru$ systems.
This is the simplest version of the QRPA, in which there is only one
intermediate state for each $J^{\pi}$.

In the SMM  the $0{\nu}$ and $2{\nu}$ moments for the $0^+\rightarrow 0^+$
transitions read  \cite{krm3}
\begin{eqnarray}
{\cal M}_{2\nu}&=& {\cal M}_{2\nu}^0  \left(\frac{\omega^{0}}
{{\omega}_{1^{+}}}
\right)^{2}\, \left(1+\,\frac{{\it G}(1^{+})} {{\omega}^{0}}\right),
\label{1}\\
{\cal M}_{0\nu} (J^{+})&=& {\cal M}_{0\nu}^0 (J^{+})\,
 \frac{\omega^{0}} {{\omega}_{J^{+}}}\,
\left(1+\,\frac{{\it G}(J^{+})} {{\omega}^{0}}\right),
\label{2}
\end{eqnarray}
where ${\cal M}_{2\nu}^0$ and ${\cal M}_{0\nu}^0 (J^{+})$ are the corresponding
unperturbed matrix elements. Here  ${\it G}(J^{+})\equiv{\it G}(pn,pn;J^{+})$
are the PP matrix elements, $\omega^{0}$ is the unperturbed energy, and
${\omega}_{J^{+}}$ are the perturbed energies. I will assume that the isospin
symmetry is strictly conserved, in which case ${\cal M}_{0\nu}(0^{+})\equiv 0$.
This statement is also valid for full calculations and therefore no further
reference will be made to the intermediate states $J^{\pi}=0^{+}$.
When the pairing factors are estimated in the usual manner, one gets
\widetext
\begin{equation}
{\omega}={\omega}^0\sqrt{1+{\it F}(34+9{\it F}/{\omega}^0)/25{\omega}^0+
16{\it G}(1+{\it F}/{\omega}^0)/25{\omega}^0},
\label{3}
\end{equation}
and
\begin{equation}
{\omega}={\omega}^0\sqrt{1+4{\it F}(45+{\it F}/{\omega}^0)/225{\omega}^0+
{\it G}(270+172{\it F}/{\omega}^0+49{\it G}/{\omega}^0)/225{\omega}^0},
\label{4}
\end{equation}
\narrowtext
for the single pair configurations $[0f_{7/2}(n)0f_{7/2}(p)]_{J^+}$ in
$^{48}Ca$
and $[0g_{7/2}(n)0g_{9/2}(p)]_{J^+}$ in $^{100}Mo$, respectively. Therefore,
while the numerators in Eq. (\ref{2}) depend only on the PP matrix elements,
their denominators depend on the particle-hole (PH) matrix elements
${\it F}(J^{+})\equiv{\it F}(pn,pn;J^{+})$, as well.
The numbers in the last two equations arise from the pairing factors.
As illustrated in Fig.\ \ref{fig2}, the SMM is a fair first-order
approximation for the $2\nu \beta \beta$ decays in $^{48}Ca$ and
$^{100}Mo$ nuclei.

The role played by the ground state correlations (GSC)
in building up Eqs. (\ref{1}) and (\ref{2}) can be summarized as follows:\\
(a) The numerator, i.e., the factor $(1+{\it G}/{\omega}^0)$, comes from the
interference between the forward and backward going contributions.
These contribute coherently in the PP channel and totally out of phase in
the PH channel.\\
(b) The ${\it G}^2$ and ${\it F}^2$ terms in the denominator are
very strongly quenched by the GSC, while the  ${\it GF}$ term is enhanced
by the same effect. In particular, for $^{48}Ca$ the term quadratic in
${\it G}$  does not contribute at all.\\
It can be stated therefore that, within the SMM and because of the GSC,
the $2\nu$ matrix element is mainly a bilinear function of ${\it G}(1^{+})$.
Besides, it passes through zero at ${\it G}(1^{+})=-{\omega}^0$ and  has a
pole when ${\omega}_{1^{+}}=0$.
Similarly, all ${\cal M}_{0\nu}(J^{+})$ moments turn out to be quotients of
a linear function of ${\it G}(J^{+})$ and the square root of another linear
function of ${\it G}(J^{+})$. Both the zero and the pole of
${\cal M}_{0\nu}(1^{+})$ matrix element coincide with those of the $2\nu$
moment. One also should  bear in mind that the magnitudes of the interaction
matrix elements ${\it G}(J)$ and ${\it F}(J)$ decrease fairly rapidly when J
increases. Thus the quenching effect, induced by the PP interaction, mainly
concerns the allowed $0\nu$ moment. For higher order multipoles it
could be reasonable to expand the denominator in Eq. (\ref{2}) in powers of
${\it G}(J^{+})/{\omega}^0$ and to keep only the linear term. This term
strongly cancels with a similar term in the numerator and the net result is a
weak linear dependence of the  ${\cal M}_{0\nu}(J^{+}\neq 1^{+})$ moments on
the PP strength. Obviously, for the last approximation to be valid, the
parameter  $t$ (or $g^{pp}$) has to be small enough to keep ${\omega}_{1^{+}}$
real.  Briefly, the SMM  can account for all four points raised
above, and leads to the following approximations for the dependence of the
$\beta \beta$ amplitudes on the PP strength
\begin{equation}
{\cal M}_{2\nu}\cong {\cal M}_{2\nu}(t=0)\frac{1-t/t_0}{1-t/t_1},
\label{5}
\end{equation}
and
\begin{eqnarray}
{\cal M}_{0\nu}&\cong& {\cal M}_{0\nu}(J^{\pi}= 1^{+};t=0)
\frac{1-t/t_0}{\sqrt{1-t/t_1}}\nonumber\\
&+&{\cal M}_{0\nu}(J^{\pi}\neq 1^{+};t=0)(1-t/t_2),
\label{6}
\end{eqnarray}
where $t_1 \geq t_0$ and $t_2 \gg t_1$, and the condition
$t\leq t_1$ is fulfilled.
It is self evident that these formulae do not depend on the type of residual
interaction, and that analogous expressions are obtained for the $\beta \beta$
matrix elements when the parameter $g^{pp}$ is used (with $g^{pp}$'s for t's).

The common behavior of the $\beta \beta$ moments for all nuclei, together
with the similarity between the SMM and the full calculations for
$^{48}Ca$ and$^{100}Mo$ (shown in Figs.\ \ref{1} and \ref{fig2},
respectively), suggests to go a step further and try  to express the
exact calculations within the framework of Eqs. (\ref{5}) and (\ref{6}).
At a first glance this seems a difficult task, because:
(i) the SMM does not include the effect of the spin-orbit splitting, which
plays a very important role in the $\beta \beta$-decay through the dynamical
breaking of the SU(4) symmetry, and (ii) the full calculations involve a
rather large configuration space (of the order of 50 basis vectors).
However, the reliability of formulae (\ref{5}) and (\ref{6}) is surprising.
The results are presented in Table \ref{tab1}.
In the upper, middle, and lower panels I show the values of the parameters
$t_0$, $t_1$, and $t_2$ that fit the  $\beta \beta$ moments displayed in the
same order in Fig.\ \ref{fig1}. I also list the values of the moments ${\cal
M}_{2\nu}$, ${\cal M}_{0\nu}(J^{\pi}= 1^{+})$, and ${\cal M}_{0\nu}(J^{+}\neq
1^{+})$ for $t=0$, together with the quantity ${\cal N}=\sqrt{{\sum}_{t=0}
[{\cal M}_{exact}(t)-{\cal M}_{fit}(t)]^2}$ that is an index of the goodness of
the fit.  The largest error occurs for $^{100}Mo$. Still, even here it is not
possible to distinguish visually the exact curves from the fitted ones. (This
makes needless the exhibition of the adjusted curves.) In fact, for this
nucleus the proposed formulae reproduce better the exact $\beta \beta$ moments
than those obtained from the SMM. It is also gratifying that all three fits
yield quite similar values for $t_0$ and $t_1$. The differences are at most of
the order of 10\%.

A comment regarding the full QRPA calculations  might be appropriate.
The  matrix element ${\cal M}_{2\nu}$ can always be expressed by the ratio
of two polynomials in ${\it G}(1^{+})$ and ${\it F}(1^{+})$ (see Eq. (8) of
Ref. \cite{krm2}). For a n dimensional configuration space these
polynomials  are of degrees 2n-1 and 2n, respectively. The above results seem
to indicate that cancellations of the type (a) and (b) are likely to be
operative to all orders, and that the linear terms in ${\it G}(1^{+})$ are
again the dominant ones.
General expressions for the $0\nu$ moments, as a function of the PP and PH
matrix elements, are not known, but a similar cancellation may be taking
place in these as well.

In summary, I have designed the Eqs. (\ref{5}) and (\ref{6}) and verified
that they nicely reproduce the full calculations of the  $\beta \beta$
matrix elements evaluated with a zero range force.
I also feel that they are of general validity, and that any modification to
the nuclear hamiltonian or to the configuration space can only change the
coefficients in these formulae, but will not lead to a different functional
dependence.
Thus, we possess now a global understanding of the $\beta \beta$ transition
mechanism (and a full control of the calculations) within the QRPA, which was
the aim of this letter.


It should be stressed that for practical application one always has to
perform the complete calculation in order to do the fit. The real advantages
of the analytic formulas (\ref{5}) and (\ref{6}) are: \\
1) they exhibit, in a very simple way, the main physics of the
$\beta \beta$-decay in the QRPA model, and summarize the common features of
the calculations done until now, and\\
2) they establish the potential and limits of the QRPA method, and give a
hint of direction that should follow the future theoretical studies.

The pole at $t=t_1$ is the response of the QRPA to the nonphysical situation,
in which the energy of the lowest virtual $J^{\pi}=1^+$ state becomes
$\cong (E_i+E_f)/2$, where $E_i$ and $E_f$ are, respectively, the energies of
the initial and final states. There is no reason in principle why this should
not happen in a nuclear model calculation (for a sufficiently large value of
t). But, within the QRPA approach the pole develops close to the "natural"
value of t, which makes the $\beta \beta$ moments to vary rather abruptly in
the physically relevant interval  $t_0 \agt t \agt t_1$. Certainly, this is a
weak point of the QRPA  \cite{hax} and it is not clear yet how it could be
circumvented.

A qualitative agreement, between the shell model and QRPA results for the
$2\nu \beta \beta$ matrix elements in $^{48}Ca$, has been reported
\cite{eng,mut}.
When applied to medium and heavy nuclei, the shell model is always accompanied
by a very severe truncation of the configuration space, in order to become
tractable. Contrarily, the QRPA is a readily accessible and fully controlled
approach, and as such it calls for further developments. Efforts in this
direction have recently been done by extending the model to describe the
$2\nu$ decays to an excited final state \cite{gri}, and by including the core
polarization corrections to the effective interaction \cite{sta}.
\acknowledgments
This research was supported by the CONICET, Argentina.
I would like to thank S. Shelly Sharma for fruitful discussions and
A.L. Plastino for a critical reading of the manuscript.

\newpage
\figure{Calculated double beta decay matrix elements ${\cal M}_{2\nu}$
(in units of $[MeV]^{-1}$),  ${\cal M}_{0\nu}(J^{\pi}=1^{+})$ and
${\cal M}_{0\nu}$, as a function of the particle-particle $S=1$,
$T=0$ coupling constant t.
The $^{48}Ca$ nucleus has been evaluated within $2\hbar \omega$ and
$3\hbar \omega$ major oscillator shells. For the remaining systems I have
adopted the oscillator shells $3\hbar \omega$ and $4\hbar \omega$ plus the
$0h_{9/2}$ and  $0h_{7/2}$ intruder orbitals from the $5\hbar \omega$ shell.
The "physical values" of the parameter t ($t_{sym}$) are shown in the
last row of Table \ref{tab1}.
\label{fig1}}
\figure{The exact (solid lines) and SMM (dashed lines) matrix elements
${\cal M}_{2\nu}$ (in units of $[MeV]^{-1}$),
as a function of the  coupling constant $t/t_0$ (defined in the text).
\label{fig2}}
\newpage
\widetext
\begin{table}
\caption{The coefficients $t_0$, $t_1$, and $t_2$ and the matrix elements
${\cal M}_{2\nu}$, ${\cal M}_{0\nu} (J^{\pi}= 1^{+})$, and
${\cal M}_{0\nu} (J^{\pi}\neq 1^{+})$ for $t=0$, in the parametrization of
the $2\nu$ and $0\nu$ $\beta \beta $ moments.
The quantity ${\cal N}$ is the norm of the residuals, i.e., the square root
of the sum of squares of the residuals.
The exact and fitted  matrix elements are equal at $t=0$, and the strength t
is varied, by steps of $\Delta t=0.1$, up to the collapse of the QRPA.
The matrix elements ${\cal M}_{2\nu}$ are given in units of $[MeV]^{-1}$.
The values of the PP coupling strength, which lead to maximal
restoration of the SU(4) symmetry ($t=t_{sym}$), are shown in the last row.}

\squeezetable
\begin{tabular}{c|ccccccc}
&{$^{48}Ca$}&{$^{76}Ge$}&{$^{82}Se$}&{$^{90}Mo$}&{$^{128}Te$}&{$^{130}Te$}\\
\tableline
$-{\cal M}_{2\nu}$ &0.173&0.308&0.321&0.451&0.381&0.331\\
$t_0$    &1.394& 1.161&1.206 &1.469 &1.265 &1.261 \\
$t_1$    &1.754& 1.680&1.691 &1.649 &2.131 &2.268 \\
${\cal N}$  &$3.26\times 10^{-2}$&$1.08\times 10^{-3}$&$7.44\times 10^{-5}$
            &$1.04\times 10^{-2}$&$2.31\times 10^{-3}$&$7.06\times 10^{-3}$\\
\tableline
$-{\cal M}_{0\nu} (J^{\pi}= 1^{+})$ &1.506&4.242&4.179&5.015&4.599&4.182\\
$t_0$     &1.244& 1.230&1.211 &1.346 &1.407 &1.408 \\
$t_1$     &1.765& 1.693&1.720 &1.741 &2.228 &2.364 \\
${\cal N}$  &$1.12\times 10^{-2}$&$4.87\times 10^{-3}$&$3.21\times 10^{-2}$
            &$2.21\times 10^{-1}$&$2.37\times 10^{-2}$&$6.34\times 10^{-2}$\\
\tableline
$-{\cal M}_{0\nu} (J^{\pi}\neq 1^{+})$ &1.501&6.924&7.495&9.762&7.997&7.486\\
$t_0$     &1.227& 1.155&1.141 &1.372 &1.377 &1.407 \\
$t_1$     &1.768& 1.741&1.764 &1.711 &2.236 &2.345 \\
$t_2$     &12.82& 13.23&12.14 &6.527 &13.39 &11.08 \\
${\cal N}$&$1.92\times 10^{-2}$&$2.46\times 10^{-2}$&$2.20\times 10^{-2}$
          &$1.11\times 10^{-1}$&$1.68\times 10^{-2}$&$3.50\times 10^{-2}$\\
\tableline
$t_{sym}$     &$\cong 1.50$&$\cong 1.25$&$\cong 1.30$&$\cong 1.50$
&$\cong 1.40$&$\cong 1.40$ \\
\end{tabular}
\label{tab1}
\end{table}
\end{document}